\documentclass[aps,prl,twocolumn,showpacs,superscriptaddress]{revtex4}
\usepackage{graphicx}
\usepackage{mathrsfs}
\usepackage{amsmath,amssymb}

\let\veps=\varepsilon
\begin{document}
 \title{Josephson current through a Kondo molecule}
\author{Rosa L\'opez}
\affiliation{Departament de F\'{i}sica, Universitat de les Illes Balears,
  E-07122 Palma de Mallorca, Spain}
\author{Manh-Soo Choi}
\affiliation{Department of Physics, Korea University, Seoul 136-701,
  Korea}
\affiliation{Department of Physics and Astronomy, University of Basel,
  Klingelbergstrasse 82, 4056 Basel, Switzerland}
\author{Ram\'on Aguado}
\affiliation{Teor\'{\i}a de la Materia Condensada,
Instituto de Ciencia de Materiales de Madrid (CSIC)
Cantoblanco,28049
Madrid, Spain.}
\date{\today}
\begin{abstract}
We investigate transport of Cooper pairs through a double quantum
dot (DQD) in the Kondo regime and coupled to superconducting
leads. Within the non-perturbative slave boson mean-field theory
we evaluate the Josephson current for two different
configurations, the DQD coupled \emph{in parallel} and \emph{in
series} to the leads. We find striking differences between these
configurations in the supercurrent as a function of the ratio
$t/\Gamma$, where $t$ is the interdot coupling and $\Gamma$ is the
coupling to the leads: the critical current $I_c$ decreases
monotonously with $t/\Gamma$ for the parallel configuration
whereas $I_c$ exhibits a maximum at $t/\Gamma=1$ in the serial
case. These results demonstrate that a variation of the ratio
$t/\Gamma$ enables to control the flow of supercurrent through the
Kondo resonance of the DQD.
\end{abstract}

\maketitle

\let\up=\uparrow%
\let\down=\downarrow%
\let\eps=\epsilon%
\let\veps=\varepsilon%
\newcommand\orbit{\mathrm{orb}}%
\newcommand\nm{\,\mathrm{nm}}%
\newcommand\K{\,\mathrm{K}}%
\newcommand\eV{\,\mathrm{eV}}%
\newcommand\mK{\,\mathrm{mK}}%
\newcommand\meV{\,\mathrm{meV}}%
\newcommand\varH{\,\mathscr{H}}%
\newcommand\half{\frac{1}{2}}%

\emph{Introduction}.--- A localized spin (magnetic impurity) forms a
many-body singlet with the itinerant electron spins surrounding it, and
hence is screened.  This, known as Kondo effect, happens at temperatures
$T$ lower than the Kondo temperature $T_K$, which is the binding energy
of the many-body singlet state~\cite{hew93}.  It manifests itself as a
quasi-particle resonance, namely, a peak of width $T_K$ at the local
density of states (LDOS) at the impurity site.  In mesoscopic structures,
such as a quantum dot (QD) coupled to normal electrodes, it leads to the
unitary conductance as predicted theoretically~\cite{theory} and
demonstrated experimentally~\cite{experiment}.
Ever since, nanoscale systems have been of particular interest as they
allow ultimate tunability and enable thorough investigations of the
Kondo physics.

It is more intriguing when one couples localized spins to
superconducting reservoirs and brings about the competition between
superconductivity and Kondo effect~\cite{Bui02}.
The subgap transport through the Andreev bound
states~\cite{And64,nato97} (ABSs) in nanostructures between
superconducting contacts is strongly affected by Kondo physics, leading
to the phase transition in the sign of Josephson
current~\cite{super-kondo}.
The large Coulomb interaction prevents the tunneling of Cooper pairs
into QD; electrons in each pair tunnel one by one via virtual
processes~\cite{super-magnetic}. Due to Fermi statistics, it results in
a negative supercurrent (i.e., a $\pi$-junction). However, this argument
is only valid in the weak coupling limit, when the gap $\Delta$ is
larger than $T_K$. In the opposite strong coupling limit ($\Delta/T_K\ll
1$) the Kondo resonance restores the positive Josephson
current~\cite{super-kondo}.  Lately, a S-QD-S system (S denotes
superconductor) has been experimentally realized, confirming some of
these physical aspects~\cite{Bui02}.

The ongoing progress in semiconductor nanotechnology has brought more
complicated nanostructures such as molecule-like double quantum
dots~\cite{dqdreview} (DQD's). The latter is an unique tunable system
for studying interactions between localized spins~\cite{dqdkondo}.
On one hand, they are basic building blocks for quantum information
processing\cite{Loss98a} as a source of entangled
states~\cite{Craig}. On the other hand, they are a prototype of the
tunable two-impurity Kondo model~\cite{Jay81}, as demonstrated
experimentally~\cite{chan,Craig}.
A natural step forward is to study the artificial molecule coupled to
superconductors. Choi, Bruder, and Loss~\cite{Mahn00} studied the
spin-dependent Josephson current through DQD in parallel in the Coulomb
blockade (CB) regime and generation of entanglement between the dot
spins.  Here, we examine the strong coupling regime, \emph{both} in
parallel and in series.
We find striking differences between these configurations in the
supercurrent as a function of the ratio $t/\Gamma$: the critical
current $I_c$ decreases monotonously with $t/\Gamma$ for the
parallel configuration whereas $I_c$ exhibits a maximum at
$t/\Gamma=1$ in the serial case.

\emph{Model---} A superconducting Kondo molecule is modelled as a
two-impurity Anderson model where the normal leads are replaced by
standard BCS s-wave superconductors. The Hamiltonian reads
\begin{widetext}
\begin{multline}
\label{hamiltonian}
\varH  = \sum_{k\alpha\in\{L,R\}\sigma}
\xi_{k\alpha}c_{k\alpha\sigma}^{\dagger}c_{k\alpha\sigma} +
\sum_{k\alpha} \left \{ \Delta_{\alpha}\exp^{i\phi_\alpha}
  c_{k\alpha\uparrow} c_{-k\alpha\downarrow} + h.c.\right \}
\\\mbox{}
+ \sum_{i\in\{1,2\}\sigma}(\varepsilon_{i\sigma}n_{\sigma}+
U_{i}n_{i\sigma}n_{i\bar{\sigma}})+ U_{12} n_{1} n_{2}+ t \sum_{\sigma} \left
  \{d_{1\sigma}^\dagger d_{2\sigma}+ h.c \right \}
+\sum_{i,k\alpha,\sigma }
\left \{ V_{i,k\alpha} c_{k\alpha\sigma}^{\dagger} d_{i\sigma}+h.c\right \} \,.
\end{multline}
\end{widetext}
$c_{k\alpha\sigma}$ ($d_{i\sigma}$) describes the electrons with
momentum $k$ and spin $\sigma$ in the lead $\alpha$ (the dot $i$);
$n_{i\sigma}\equiv d_{i\sigma}^\dag d_{i\sigma}$.  $\phi_{\alpha}$
($\Delta_{\alpha}$) denotes the superconducting phase (gap) for the lead
$\alpha$. $t$ ($V_{ik\alpha}$) is the tunneling amplitude between the
dots (the dots and leads).  For a DQD in series $V_{1,kR}=V_{2,kL}=0$
with $V_{1,kL}=V_{2,kR}=V_0$; for a DQD in parallel
$V_{1,kR}=V_{1,kL}=V_{2,kL}=V_{2,kR}=V_{0}$.  $\veps_i$ are the
single-particle energies on the dots tuned by gate voltages.  $U_i$
is the on-site Coulomb interaction on the $i$th dot, and $U_{12}$ is the
interdot Coulomb interaction.

We will be interested in the limit $U_1,U_2\to\infty$, $U_{12}=0$, and
$-\veps_i\gg\Gamma$ so that $\langle{n_{i\sigma}}\rangle=1$ for each
$i=1,2$.
The model in this limit is well described in the slave-boson
language~\cite{sbmft1}.  We first write the physical fermionic
operator as $d_{i,\sigma}=b^\dagger_if_{i,\sigma}$, where
$f_{i,\sigma}(b^\dagger_i)$ is the pseudofermion(boson) which
destroys(creates) one ``occupied (empty) state'' on the dot $i$. We then
introduce two constraints which prevent double occupancy in each dot
($U_1,U_2\to\infty$) by means of Lagrange multipliers, $\lambda_1$ and
$\lambda_2$. 
The resulting model is solved within the mean-field (MF) approach, namely,
replacing
\begin{math}
b_i(t) \rightarrow
\langle b_i \rangle
\equiv \tilde{b}_i
\end{math}
and thereby neglecting charge fluctuations in the dots.
This approach has been applied successfully to the Kondo molecules
coupled to normal leads~\cite{dqdkondo}.

\begin{figure}
\centering %
\includegraphics*[width=70mm]{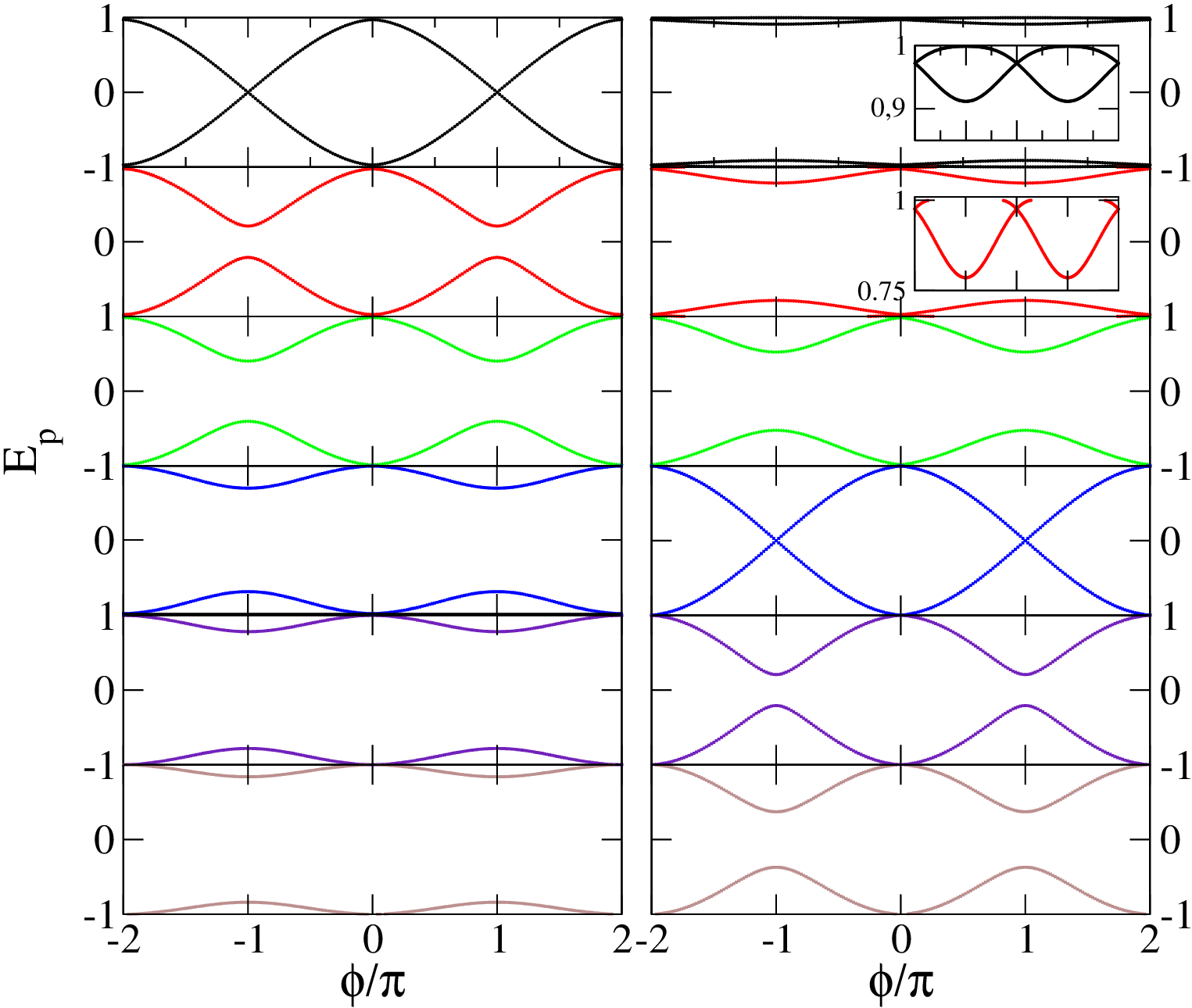}
\caption{(Color online). Andreev bound states in units of $\Delta$.
  Left panel: Parallel case for
  $t/\Gamma=0,0.25,0.5,1,1.25,1.5$ (from top to bottom).
  Right panel: Serial case for $t/\Gamma=0.1,0.25,0.5,1,1.25,1.5$
  (from top to bottom).  In both cases, $\Delta=0.1T_K^0$.}
\label{fig1}
\end{figure}

\begin{figure}
\centering %
\includegraphics*[width=75mm]{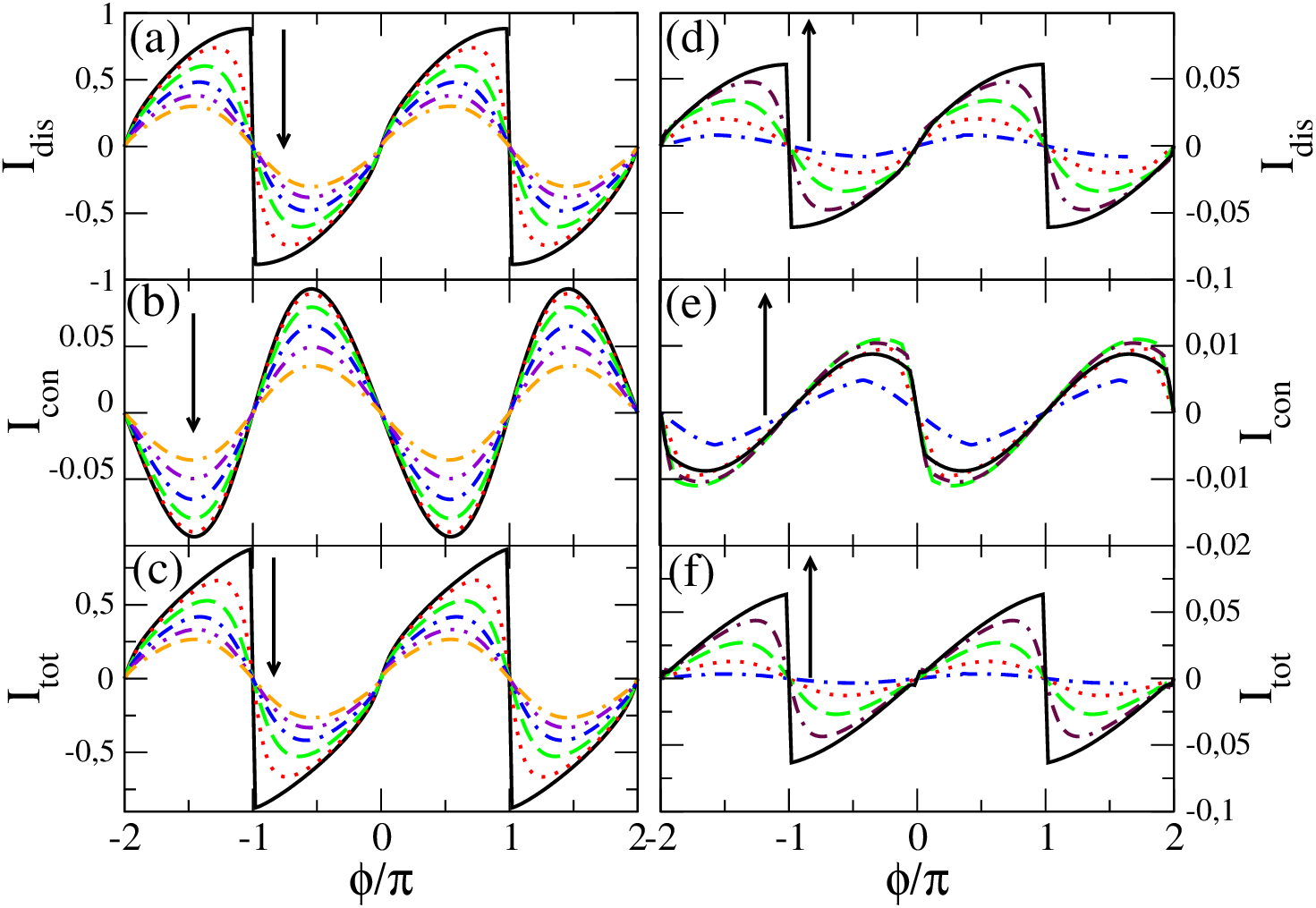}
\caption{(Color online). $t/\Gamma\leq 1$. Left(Right) panel corresponds to
  the parallel(serial) case. [a(d)] discrete $I_{\rm dis}$, [b(e)] continuum
  $I_{\rm con}$ and [c(f)] total supercurrent $I_{\rm tot}$ versus $\phi$ for
  $t/\Gamma=0,0.2,0.4,0.6,0.8,1$ ($t/\Gamma=0.2,0.4,0.6,0.8,1$) and
  $\Delta=0.1 T_{K}^0$.  Currents are in units of $2|e|/\Delta\hbar$.
  The curves are arranged with increasing $t/\Gamma$ in the direction of
  the arrows.}
\label{fig2}
\end{figure}

\begin{figure}
\includegraphics*[width=75mm]{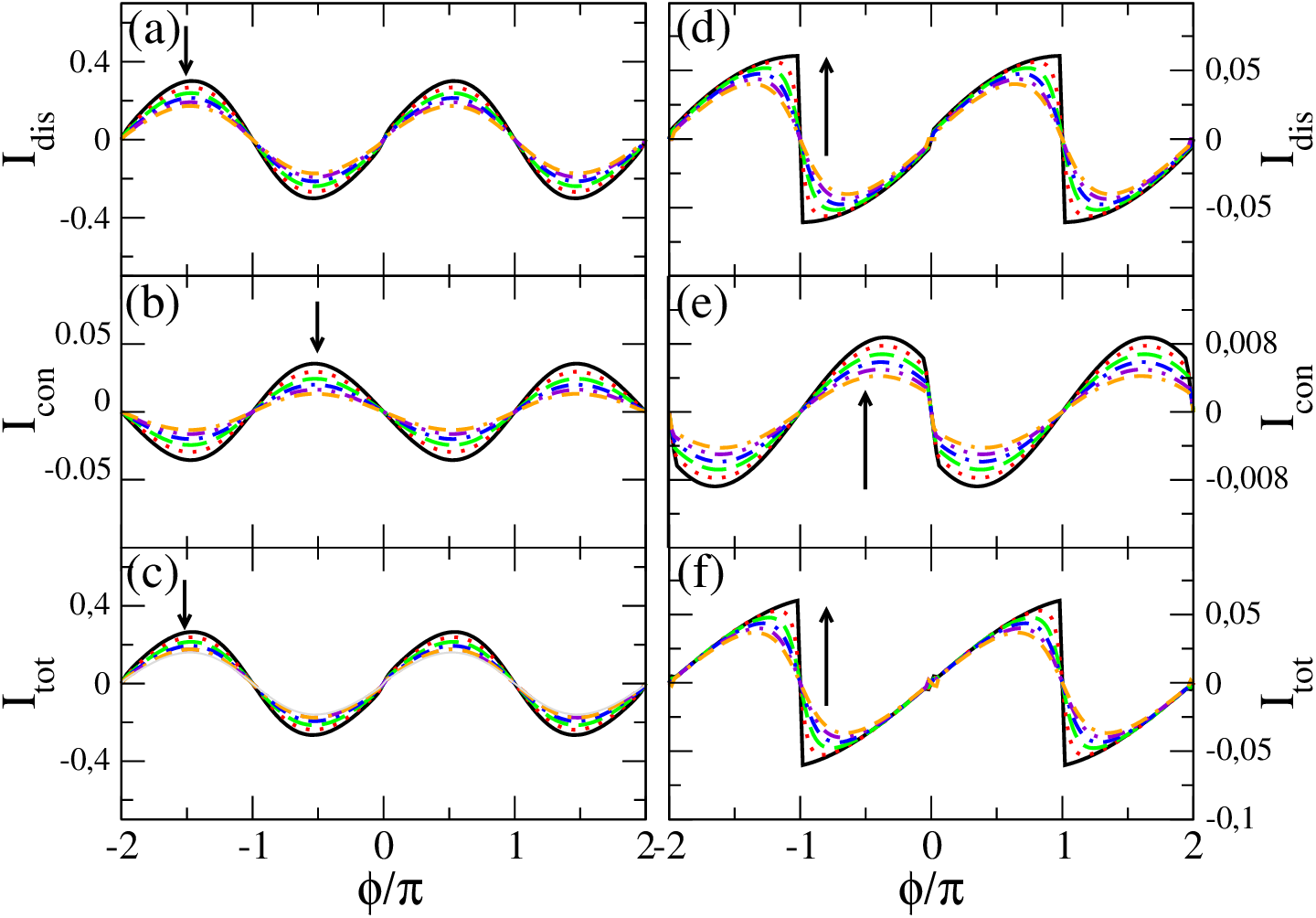}
\caption{(Color online). The same as Fig.~\ref{fig2} except that
  $t/\Gamma\geq 1$.}
\label{fig3}
\end{figure}

The MF version of $\varH$ is now quadratic and contains four
parameters, i.e., $\tilde{b}_{1,2}$ that renormalizes the tunneling amplitudes, $\tilde{V}_{i,k\alpha}=\tilde{b}_{i}V_{i,k\alpha}$ and $\tilde t= t b_{1} b_{2}$ and
$\lambda_{1(2)}$ that renormalizes the energy levels
$\tilde{\varepsilon}_{1,2\sigma}=\varepsilon_{i}+\lambda_{1,2}$.  They are
determined from the solution of the MF equations in a
self-consistently fashion. The MF equations become simpler in the
\emph{Nambu-Keldysh} space where $\Psi_{k\alpha\sigma}^\dagger=\left(c_{k\alpha\sigma}^{\dagger},  c_{-k\alpha\bar{\sigma}}\right)$ and $\Phi_{i\sigma}^\dagger=\left(f_{i\sigma}^\dagger, f_{i\bar{\sigma}}\right)$ are the spinors for the
conduction and localized electrons. From the equation-of-motion (EOM) of the
boson fields and the constrains the MF equations read:
\begin{subequations}\label{meanfieldeq}
\begin{eqnarray}
\label{eq1}
&& \tilde{b}_{1(2)}^2+ \frac{1}{N}\sum_{\sigma} \langle
\Phi_{1(2)\sigma}^\dagger(t) \hat{\sigma}_z\Phi_{1(2)\sigma}(t)\rangle=\frac{1}{N}\,,
\\
\label{eq2}
&&\frac{1}{N} \sum_{\alpha\in\{L,R\}\sigma} \tilde{V}_{1(2),k\alpha} \langle
 \Psi_{k\alpha\sigma}^\dagger (t)\hat{\sigma}_z
 \Phi_{1(2)\sigma}(t)\rangle\,
\\
 &&
+\frac{\tilde{t}}{N} \sum_\sigma\langle \Phi_{1(2)\sigma}(t) \hat{\sigma}_z
\Phi_{2(1)\sigma}(t) \rangle +\lambda_{1(2)}\tilde{b}_{1(2)}^2=0\,,\nonumber
\end{eqnarray}
\end{subequations}
Hereforth, for simplicity we assume a symmetric structure with
$\veps_{1(2)\sigma}=\veps_{0\sigma}$, $\Delta_{L(R)}=\Delta$, and
$\phi_{L}=-\phi_R=\phi$. The system of MF equations (\ref{meanfieldeq})
is now written in terms of the $2\times2$ matrix lesser Green function
(GF) for the dots $\hat
G_{i,j\sigma} (t,t')= i\langle
\Phi_{j\sigma}^\dagger(t'),\Phi_{i\sigma}(t)\rangle$ and the lead-dot
matrix lesser GF $\hat G_{i,k\alpha\sigma} (t,t')= i\langle
\Psi_{k\alpha\sigma}^\dagger(t'),\Phi_{i\sigma}(t)\rangle$ (with components $\hat G^{11}=G$, ``electron-like'' GF $\hat G^{12}=F$
is the anomalous propagator, $\hat G^{21}=F^\dagger$, and $\hat
G^{11}=\tilde{G}$ corresponds to the ``hole-like'' propagator). Following the standard procedure, the lesser GF's are obtained applying
rules of analitical continuation along a complex time contour to the EOM
of the time-ordered GF (for details, see Ref.~\cite{sbmft2}). The two
off-diagonal
GF's, namely, $\hat G_{i,j\sigma}$ (with $i\neq j$) lead-dot GF $\hat G_{i,k\alpha \sigma}$ can be cast in terms of
the diagonal dot GF $\hat G_{i,i\sigma}\equiv\hat G_{i\sigma}$ using
the EOM technique.  By doing this Eq.~(\ref{meanfieldeq}) becomes (for simplicity we omit the spin indices)
\begin{subequations}
\label{mfeqs}
\begin{eqnarray}
\label{mfeq1}
&& \frac{\tilde{\Gamma}}{\Gamma}+ \int_{-\infty}^{\infty}
\frac{d\epsilon}{2\pi i}\left [G_{1(2)}^{<}(\epsilon)+\tilde{G}_{1(2)}^{<}(\epsilon)\right]=0\,,
\\
&&\frac{\tilde{\Gamma}}{\Gamma} (\tilde{\veps}_{1(2)}-\veps_{1(2)})+ \int_{-\infty}^{\infty}
\frac{d\epsilon}{2\pi i}
  [G_{1(2)}^{<}(\epsilon)(\epsilon-\tilde{\veps}_{1(2)})\nonumber
\\
\label{mfeq2}
&&+\tilde{G}_{1(2)}^{<}(\epsilon)(\epsilon+\tilde{\veps}_{1(2)})]=0\,,
\end{eqnarray}
\end{subequations}
where $\tilde{\veps}_i=\veps_0+\lambda_{i}$
are the renormalized levels and $\tilde{\Gamma}_{i}=\tilde{b}_{i}^2\Gamma$
the renormalized hybridization, which are equal to the Kondo
temperature $T_K$ for the coupled system~\cite{noteT_K}
[$\Gamma=2\pi\rho_N V_{0}^2$ with $\rho_N$ being the normal-state DOS].
At equilibrium we can employ $\hat{G}_{i}^{<}=2if(\epsilon)\mathrm{Im}
\hat{G}_{i}^{r}$ with
$f(\epsilon)$ being the Fermi function.
$\hat{G}_{i}^{r}$ is determined by the lead-dot
$\hat{\Sigma}_{i\alpha}^{r}$ and the interdot tunneling
$\hat{\Sigma}_{ti}^{r}$ self-energies.
Thus, the matrix elements of the lead-dot self-energy are:
$\hat{\Sigma}_{i\alpha}^{r,11(22)}= - i\tilde{\Gamma}_{\alpha}^{i}\rho_{S}(\epsilon)$,
$\hat{\Sigma}_{i\alpha}^{<,12}=(\hat{\Sigma}_{i\alpha}^{<,21})^*=-i\tilde{\Gamma}_{\alpha}^{i}\rho_{S}(\epsilon)\Delta\exp(i\phi_{\alpha})/|\epsilon|$
with $\rho_{S}(\epsilon)=|\epsilon|/\sqrt{\epsilon^2-\Delta^2}\theta(\epsilon-\Delta)$
as the superconducting DOS and $\tilde{\Gamma}_{\alpha}^{i}=2\pi
|\tilde{V}_{ik\alpha}|^2\rho_N$. The interdot tunneling self-energy is $\hat{\Sigma}_{t1(2)}^r=\tilde{t}^2
\hat{\sigma}_z [g_{2(1)}^{r}-\sum_{\alpha}
  \hat{\Sigma}_{2(1)\alpha}^{r}] \hat{\sigma}_z
$, where ${\sigma}_z$ is the $z$ component of the Pauli matrices and
$\hat{g}_{i}^{r}$ is the matrix GF for an isolated QD. Since we deal with a
symmetric structure
$\lambda_{i}=\lambda\rightarrow\tilde{\veps_{i}}=\tilde{\veps}_{0}$ and
$\tilde{\Gamma}=\tilde{\Gamma}_{i}$. Thus, for example for the serial configuration the diagonal components of the GF read
\begin{equation}
\label{super::eq:1}
G^{r(a)} =
\frac{[(A\pm\tilde{\veps}_0)\left[(A^2-\tilde{\veps}_0^2)-s(\epsilon)^2+\tilde{t}^2
    (A\mp\tilde{\veps}_0)\right]}{\mathcal{D}(\epsilon)}\,,
\end{equation}
where
\begin{math}
\mathcal{D}(\epsilon)
= \left[A^2-X^2-Y^2-(\tilde{\veps}_0-\tilde{t})^2\right]^2
-4\tilde{t}^2(Y^2+\tilde{\veps}^2)
\end{math}
with $X=s(\epsilon)\cos(\phi)$, $Y=s(\epsilon)\sin(\phi)$,
$A=\epsilon\left[1+s(\epsilon)/\Delta\right]$, and
$s(\epsilon)=\Delta\tilde{\Gamma}/\sqrt{\Delta^2-\epsilon^2}$.
The GF~(\ref{super::eq:1}) describes the discrete Andreev bound
states in the subgap region ($|\epsilon|<\Delta$) as well as the
continuum spectrum above the gap ($|\epsilon|>\Delta$).  The
Andreev states appear as poles of the GF; i.e., the solutions of
$\mathcal{D}(\epsilon)=0$ (see Fig.~\ref{fig1}).  Accordingly, the
Josephson current has two contributions,
$I_\mathrm{tot}=I_\mathrm{dis}+I_\mathrm{con}$; $I_\mathrm{dis}$
from the discrete Andreev states and $I_\mathrm{con}$ from the
continuum (see Figs.~\ref{fig2} and \ref{fig3}). The two parts of
the Josephson current for the DQD in series (see below and
Ref.~\cite{note} for the DQD in parallel) are given by (hereafter
currents are expressed in units of $2|e|\Delta/\hbar$): 
\begin{subequations}
\label{current}
\begin{eqnarray}
\label{currentd}
I_\mathrm{dis} & = & -2 t^2 \sin(2\phi)\sum_{E_p}
\left.\frac{s^2(\epsilon)}{\Delta\mathcal{D}'(\epsilon))}\right|_{\epsilon=E_p}\,,\\
\label{currentc}
I_\mathrm{con} & = & -2 t^2
\sin(2\phi)\mathrm{Im}\int_{-\infty}^{-\Delta}{d\epsilon}\;
\frac{1}{\Delta\mathcal{D}(\epsilon)} \,.
\end{eqnarray}
\end{subequations}
In Eq.~(\ref{currentd}) the summation is over all Andreev states $E_p\in
[-\Delta,\epsilon_F]$ ($\epsilon_F$ is the Fermi energy).
Interestingly, we will see below [Eqs.~(\ref{super::eq:2}) and
(\ref{super::eq:4}) ] that in the deep Kondo limit ($\Delta\ll T_K$), we
recover the short-junction limit for the Josephson current through
non-interacting resonance levels~\cite{Bee92}.  In this limit the
continuum contribution is almost negligible. For $\Delta\lesssim T_K$
the contributions from the continuum part becomes considerable and the
behavior deviates from the non-interacting case.

Next we present our results for the DQD in parallel and in
series, respectively.
We will choose $\veps_{0}=-3.25$, $D=100$ (bandwidth), $\epsilon_F=0$
 and different values for the rest of parameters. For these
values $T_{K}^0=D\exp{-\pi|\veps_{0}|/\Gamma=0.0036}$. (All
energies are given in units of $\Gamma$).

\emph{DQD in parallel---} The problem is greatly simplified by the
transformation $c_{k\alpha\sigma}=(c_{k_e\sigma}\pm
c_{k_{o}\sigma})\sqrt{2}$ and $f_{1(2)\sigma}=(f_{e\sigma}\pm
f_{o\sigma})\sqrt{2}$.  The MF Hamiltonian is mapped into two
independent Josephson junctions (``even'' and ``odd'') through
\emph{effective} resonant levels at $\tilde{\veps}_{0}\pm\tilde{t}$.
Each of the two resonant levels accommodates an Andreev state $E_{e/o}$
and carries
Josephson current~\cite{note};
$I_\mathrm{tot}(\phi)=-2 e\Delta/\hbar \left(I_{e} + I_{o}\right)$ (see
Ref~\cite{note} for their explicit expressions). Typical profiles of
Andreev states on DQD in parallel are shown in the
left panels of Fig.~\ref{fig1}. Josephson currents are shown in the
left panels of Fig.~\ref{fig2} ($t\leq\Gamma$) and Fig.~\ref{fig3}
($t\geq\Gamma$).
We note that in the deep Kondo limit $\Delta\ll T_K$ the DQD in parallel
behaves in effect like two non-interacting resonance levels~\cite{Bee92}:
The Andreev states and the Josephson currents are
given by the corresponding expressions for the short-junction limit
$E_{e/o}=\Delta[1-\mathcal{T}_{e/o}\sin^2(\phi)]^{1/2}$ and
\begin{equation}
\label{super::eq:2}
I_{e/o}(\phi)
=\mathcal{T}_{e/o}\sin{2\phi}
\left[1-\mathcal{T}_{e/o}\sin^2{\phi}\right]^{-1/2}\,,
\end{equation}
respectively,
with $\mathcal{T}_{e/o}=4
\tilde{\Gamma}^2/(\tilde{\veps}_{e/o}^2+\tilde{\Gamma})$. For very small $t/\Gamma$ both dots have their own Kondo resonances at
$\epsilon_F$ and the Josephson current resembles that of a ballistic
junction. As $t/\Gamma$ increases the even and odd Kondo resonances
$\tilde\veps_{0}\pm \tilde{t}$ move away from $\epsilon_F$ and as a result
$I_{\rm tot}(\phi)$
diminishes becoming more sinusoidal $I_{e/o}(\phi)\approx
\mathcal{T}_{e/o}\sin{2\phi}$.

Experimentally more accessible is the critical current
$I_c$~\cite{Jar06}, which is given by $I_c(\phi)\approx 2
e\Delta/\hbar\sum_{\beta\in\{e,o\}} [1-(1-\mathcal{T_{\beta}})^{1/2}]$
and hence decreases with $t/\Gamma$.  Notice that in spite of the simple
formal expression for $I_{e/o}$, $I_c$ depends on the \emph{many-body}
parameters, $T_{K}$ and $\tilde{\veps}_{0}\pm\tilde{t}$, in a nontrivial manner
through the solution of Eq.~(\ref{mfeqs}).

\emph{DQD in series---} A completely different physical scenario is
found for the serial configuration. Here the even/odd channels are no
longer decoupled and cause novel interference.
The manifestation of the interference can be first seen in the profiles
of the Andreev states as depicted in Fig.~(\ref{fig1}).  An important
difference from the parallel case is the almost flat spectrum with
values close to $E_{e/o}(\phi)\lesssim\Delta$ (reflecting a very small
supercurrent as seen below).  As $t/\Gamma$ increases the spectrum
possesses larger amplitude, and for $t/\Gamma\approx 1$ we eventually
recover the spectrum of a ballistic junction $E_{e/o}(\phi)\approx\Delta
\cos(\phi)$. For $t/\Gamma\geq 1$ gaps are opened again (suggesting that
$I_\mathrm{tot}(\phi)$ diminishes, see below).

\begin{figure}
\centering %
\includegraphics*[width=65mm]{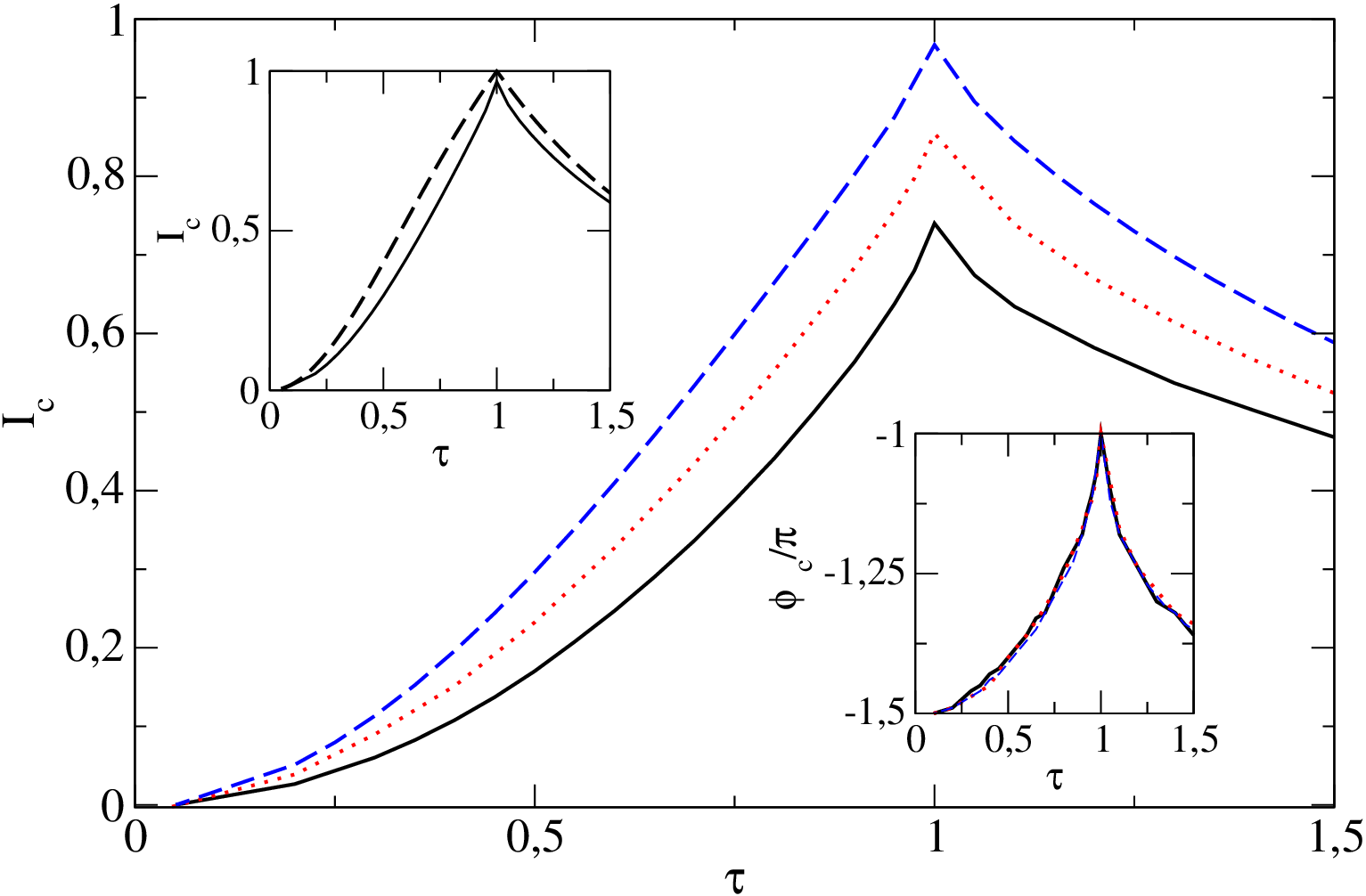}
\caption{(Color online). Critical current as a function of $t/\Gamma$
  for $\Delta/T_K^0=0.1, 0.25, 0.5$ (top to bottom). Upper left inset:
  comparison with the Josephson critical current given by Eq.~(\ref{super::eq:4}) (dashed line) for $\Delta/T_K^0=0.1$. Lower
  right inset: critical phase at which the maximum current
  occurs.}\label{fig4}
\end{figure}

Formally, the supercurrent, for $\Delta\ll T_K$, is still given by
the expression
\begin{equation}
\label{super::eq:4} I(\phi) = \mathcal{T}\sin{2\phi}
\left[1-\mathcal{T}\sin^2{\phi}\right]^{-1/2},
\end{equation}
but now the transmission $\mathcal{T}$ is
\begin{math}
\mathcal{T}=4 \tilde{t}^2
\tilde{\Gamma}/[((\tilde{\veps}_0-\tilde{t})^2+\tilde{\Gamma}^2)
((\tilde{\veps}_0+\tilde{t})^2+\tilde{\Gamma}^2)]
\end{math}. We can interpret that for small $t/\Gamma$ the Cooper pairs
hop directly between the two Kondo resonances and
$\mathcal{T}\propto t^2$. The supercurrent presents a
sinusoidal-like behavior as shown in Fig.~\ref{fig2}.  With
$t/\Gamma$ increasing, the physical situation changes drastically
around $t/\Gamma=1$, where the Kondo singularities of each dot
hybridize into a correlated state as a result of the coherent
superposition of both Kondo states. We find $\mathcal{T}\approx 1$
and consequently the supercurrent-phase relation exhibits a
typical ballistic-junction behavior (see Fig.~\ref{fig3}). Further
increasing $t/\Gamma$ makes $I_\mathrm{tot}(\phi)$ smaller, which
is attributed to the formation of bonding and antibonding Kondo
resonances. This results in a nonmonotonous behavior of $I_c$ as a
function of $t/\Gamma$ (shown in Fig.~\ref{fig4} for
$\Delta/T_K^0=0.1, 0.25, 0.5$, from top to bottom), with a maximum
at $t=\Gamma$ (coherent superposition of both Kondo resonances).
For the lowest gap (in the short junction regime) the maximum
critical current reaches the universal value of $2e\Delta/h$ as
expected~\cite{Bee92}.

The physics of the tunability of $I_c$ as a function of $t/\Gamma$
is similar to the transistor-like control of supercurrents in a
carbon nanotube quantum dot connected to superconducting
reservoirs recently reported by Jarillo {\it et
al}.,~\cite{Jar06}.  These experiments demonstrate that the
supercurrent flowing through the QD can be varied by means of a
gate voltage which tunes on- and off-resonance successive discrete
levels of the QD with respect to $\epsilon_F$ of the reservoirs.
In our case, the Kondo resonances play the role of the discrete
levels in the experiments of Ref.~\cite{Jar06} whereas $t/\Gamma$
is the extra knob that tunes the position of the Kondo resonances
and thus modulates the critical current. Interestingly, in our
case, i) the supercurrent is mediated by a coherent many-body
state (the Kondo resonance) instead of a single particle one and
ii) the maximum supercurrent at $t=\Gamma$ corresponds to coherent
transport of Cooper pairs through the whole device whereas an
increase of $t/\Gamma\geq 1$ splits the Kondo resonance into two
(bonding and antibonding) resulting in a splitting of the Cooper
pair into two electrons (one on each resonance) and thus a
reduction of the supercurrent. This suggest the use of the Kondo
effect as an alternative to previous proposals using
DQDs~\cite{Mahn00} for generating and manipulating entangled pairs
in a controlled way.

For the experimental realization of the superconducting Kondo DQD, we
propose carbon nanotubes since (i) they show Kondo
physics~\cite{cntkondo}, (ii) it is possible to fabricate tunable
double quantum dots~\cite{cntdqd}, and (iii)
they are ideal systems to attach new material as
electrodes~\cite{Bui02}.

\emph{Conclusions}.---In closing we have studied Cooper pair
transport through an artificial Kondo molecule. We find remarkable
differences in the phase-current relation when the DQD is built in
series or in parallel. For a DQD in parallel, the supercurrent
always decreases with $t/\Gamma$ whereas for a serial Kondo
molecule the current behaves nonmonotonously. This fact allows an
extra control of the critical current, and thus of Cooper pairs,
through Kondo molecules by simply tuning the interdot tunneling
coupling.

We acknowledge D.~S\'anchez for fruitful and long discussions.
This work was supported by MAT2005-07369-C03-03 and the Spanish MEC
through the program Ramon y Cajal program (R.L.),
the SRC/ERC program (R11-2000-071), the KRF Grant (KRF-2005-070-C00055),
the SK Fund, and the KIAS.


\begin{thebibliography}{10}

\bibitem{hew93} A.C.~Hewson, {\it The Kondo Problem to Heavy Fermions}
  (Cambridge University Press, Cambridge, UK, 1993)
\bibitem{theory} LI.~Glazman and M.E.~Raikh, Pis'ma Zh. Éksp. Teor. Fiz.
  {\bf 47}, 378 (1988) [JETP Lett. {\bf 47}, 452 (1988)].
T.K.~Ng and P.A.~Lee, Phys. Rev. Lett.{\bf 61}, 1768 (1988).
\bibitem{experiment} D. Goldhaber-Gordon {\it et al}., Nature {\bf 391}, 156
  (1998). S.M.~Cronenwett {\it et al}.,  Science {\bf 281}, 540 (1998).
J.~Schmid {\it et al}., Physica B {\bf 256-258}, 182 (1998).
\bibitem{Bui02}
M.R.~Buitelaar, T. Nussbaumer, C. Schonenberger, Phys. Rev. Lett. 89, 256801 (2002).
M.R.~Buitelaar {\it et al}., Phys. Rev. Lett. 91, 057005 (2003).
\bibitem{And64}
A.F.~Andreev,  Zh.~Éksp.~Teor.~Fiz.~{\bf 46},1823 (1964) [Sov.
phys. JETP {\bf 19}, 1228 (1964)]; {\it ibid} {\bf 49}, 655 (1966)
[{\bf 22}, 455 (1966)].
\bibitem{nato97}
B.J.~van Wees and H.~Takayanigi, in {\it Mesoscopic Electron
Transport}, edited by L.L.~Sch\"on {\it et al}. (Kluwer,
Dordrecht, 1997).


\bibitem{super-kondo}
G.~Sellier {\it et al}., cond-mat/0504649 (unpublished) (2005).
M.S.~Choi, M. Lee, K. Kang, and W. Belzig, Phys. Rev. B \textbf{70}, 020502(R)
(2004); {\it ibib} Phys. Rev. Lett \textbf{94}, 229701 (2005).
F.~Siano and R. Egger, Phys. Rev. Lett. \textbf{93}, 047002 (2004);
Phys. Rev. Lett \textbf{94}, 229702 (2005).
E.~Vecino, A. Martin-Rodero, and A.L. Yeyati, Phys. Rev. B \textbf{68}, 035105 (2003).
A.L.~Yeyati, A. Martin-Rodero, and E. Vecino, Phys. Rev. Lett. \textbf{91}, 266802 (2003).
Y.~Avishai, A. Golub, and A.D. Zaikin, Phys. Rev. B \textbf{67}, 041301(R) (2003).
A.V.~ Rozhkov and D.P.~Arovas, Phys. Rev. B \textbf{62}, 6687 (2000). A.V.~
Rozhkov, D.P. Arovas, F. Guinea,  Phys. Rev. B \textbf{64}, 233301 (2001).
A. A.~Clerk and V.~Ambegaokar, and S. Hershfield,  Phys. Rev. B \textbf{61}, 3555 (2000).
\bibitem{super-magnetic}
I.O-.~Kulik, Zh.~Éksp.~Teor.~Fiz. \textbf{49}, 585 (2000).[Sov. phys. JETP {\bf 22}, 841
(1966)]. H.~Shiba and T.~Soda, Prog. Teor. Phys. \textbf{41}, 25 (1969). L.I.~Glazman and K.A.~Matveev, Pis'ma Zh.~Éksp.~Teor.~Fiz.~[JETP Lett. {\bf
  49}, 659 (1989)].
B.I.~Spivak and S.A.~Kilvelson, Phys. Rev. B \textbf{43}, R3740 (1991).
\bibitem{dqdreview} For a recent review see W.G.~van der Wiel, S.~De
  Franceschi, J.M.~Elzerman, T.~Fujisawa, S.~Tarucha, and
  L.P.~Kouwenhoven, Rev. Mod. Phys. {\bf 75}, 1 (2003).
\bibitem{dqdkondo}
A.~Georges and Y.~Meir, Phys. Rev. Lett. {\bf 82}, 3508 (1999). R.~Aguado and
D.C.~Langreth, Phys. Rev. Lett. {\bf 85}, 1946 (2000), {\it ibib} Phys. Rev. B {\bf 67}, 245307 (2003). T.~Aono and M.~Eto,
Phys. Rev. B {\bf 63}, 125327 (2001). R.~Lopez, R. Aguado, and G. Platero,
Phys. Rev. Lett. {\bf 89}, 136802 (2002).
 P. Simon, R. Lopez, and Y. Oreg {\it et al}., {\bf 94}, 086602 (2005).
\bibitem{Loss98a}
  D. Loss and D.P. DiVincenzo, Phys. Rev. A \textbf{57}, 120 (1998).
\bibitem{Craig}
 N.J.~Craig {\it et al}., Science {\bf 304} 565 (2004).
\bibitem{Jay81}
 C. Jayaprakash, H. R. Krishna-murthy, and J. W. Wilkins, Phys. Rev. Lett. {\bf
 47},
 737 (1981).
\bibitem{chan} H.~Jeong {\it et al}., Science {\bf 293}, 2221
  (2001). J.C.~Chen, A.M. Chang, M.R. Melloch, Phys. Rev. Lett. \textbf{92}, 176801
  (2004).
\bibitem{Mahn00}
M.S. Choi, C. Bruder, D. Loss, Phys. Rev. B \textbf{62}, 13569 (2000).
\bibitem{sbmft1}
P.~Coleman, Phys. Rev. B {\bf 29}, 3035 (1984).
For a review, see D.M. Newns and N. Read, Adv. Phys.
\bibitem{sbmft2}
R. Lopez, R. Aguado, G. Platero, Phys. Rev. B {\bf 69}, 235305 (2004).

\bibitem{noteT_K} For a DQD connected in series to \emph{normal} leads,
  $T_K$ has an exponential dependence on $t$; $T_K\approx T_K^0 e^{
    t/\Gamma {\rm tan}^{-1}(t/\Gamma)}$.
\bibitem{note} For the DQD in parallel, the ABSs and Josephson current
  are conveniently expressed in the even/odd basis.  The ABSs are given
  as the solutions of $\mathcal{D}_{e/o}(\epsilon)=\epsilon^2
  (\sqrt{\Delta^2-\epsilon^2}+\tilde{\Gamma})^2-{(\tilde{\veps}_0\pm
    \tilde{t})^2(\Delta^2-\epsilon^2)-\Delta^2\Gamma^2\cos^2(\phi)}$. The
  discrete and continuum parts of the current are $I^{e/o}_{\rm
    dis}=-\sin(2\phi)
  s^2(\epsilon)/\Delta\mathcal{D}^{'}_{e/o}|_{\epsilon=E_{e/o}}$ and
  $I^{e/o}_{\rm con}=-\sin(2\phi)\int_{-\Delta}^{0} d\epsilon
  s^2(\epsilon) \mathrm{Im}[1/\Delta \mathcal{D}_{e/o}(\epsilon)]$,
  respectively.

\bibitem{Bee92}
C.~W.~J.~Beenakker and H.~Van Hounten. {\it Single-electron tunneling and
Mesoscopic devices}, edited by H.~Koch, and H.~Lübbig (Springer Berlin
1992).
\bibitem{Jar06}
P. Jarillo-Herrero, J. A. van Dam, and L.P.~Kouwenhoven, Nature {\bf 439}, 953
(2006).
\bibitem{cntkondo}
J.~Nyg\aa rd {\it et al}.,  Nature (London) \textbf{408}, 342 (2000).
\bibitem{cntdqd}
N.~Manson {\it et al}.,Science \textbf{303} 655 (2004). S. Sapmaz \textit{et
  al}., cond-mat/0602424 (unpublished). M. R. Graeber \textit{et al}., cond-mat/0603367 (unpublished).
\end{thebibliography}
\end{document}